\documentclass[manuscript]{aastex}

\received{}
\revised{}
\accepted{}

\shortauthors{Jones et al.}
\shorttitle{Massive Outflows}

\begin{document}

\title{OUTFLOWS FROM LUMINOUS YSOs: An Infrared Polarimetric Study}
\author{Terry J. Jones~\altaffilmark{1}, Charles E. 
Woodward~\altaffilmark{1}, and Michael S. Kelley~\altaffilmark{1}}
\affil{Department of Astronomy, University of Minnesota, 116 Church 
Street S.E., Minneapolis, MN 55455}

\altaffiltext{1}{Visiting Astronomer at the Infrared Telescope Facility which is operated by the University of Hawaii under contract from the National Aeronautics and Space Administration.}

\begin{abstract}

We present Near Infrared imaging polarimetry of three regions of massive star formation, G$192.16 - 3.82$, Cepheus A, and W42. In W42 we have discovered a new bipolar nebula located at the far side of the HII region behind the visible cluster of exciting stars. The axis of this new nebula is aligned with the magnetic field threading the entire cluster region. Polarization in the bipolar outflow nebulosity associated with G192.16 is consistent with a single illuminating source, too faint to be detected at $2~\micron$. Polarization in the reflection nebulosity associated with Ceph A requires more than one illuminating source, although HW2 is clearly dominant. In all three objects, the magnetic field in the outflow at distances greater than $\sim 0.2$ pc is radial. In G192.16 the magnetic field geometry closer than $\sim 0.2$ pc to the embedded star appears chaotic. For G192.16 the outflow is not aligned with the surrounding magnetic field, which lies in the galactic plane. In Ceph A, the outflow axis could be interpreted as being aligned with the galactic plane, but the magnetic field threading the region is not. Only in the case of W42 is the magnetic field threading the HII region aligned with the mean field in the surrounding galactic plane.

\end{abstract}


\section{INTRODUCTION}

Molecular outflows are a common occurrence in the evolution of young stellar objects (YSOs) of all masses \citep{lad85, bac96}. The driving mechanism for these bipolar outflows is still a subject of considerable debate and most of our knowledge of this phenomenon is based on the study of low mass YSOs \citep{cab97,sha98}. This is understandable because low mass YSOs are more common, closer on average, and often are less obscured by dust than deeply embedded, high mass YSOs. Currently, models invoking centrifugally driven winds are a popular mechanism for driving these outflows in low mass YSOs \citep[e.g.][]{kud03}. In these models the magnetic field is twisted into a spiral pattern rising vertically off of the star's dense, rotating circumstellar disk. Material entrained in this field is forced up the vertical axis and produces the observed collimated outflow \citep{pud97, kud97}.  \citet{kon99} reviewed efforts to extend these magnetically driven wind models of bipolar outflows in low-mass stars to high mass YSO's. It is not clear if the massive outflows associated with very luminous YSOs are simply a scaled up version of these magnetically driven winds, or a completely different process. 

Determining the magnetic field geometry in massive bipolar outflows could provide important constraints on these models, but this is a difficult observational task. These outflows are most easily studied when the outflow axis lies in the plane of the sky and the star's equatorial disk is viewed nearly edge-on. This projection produces the classic bow-tie morphology seen in many sources and allows both poles of the outflow to be easily separated spatially on the sky. If dust grains in the bipolar lobes are aligned by the magnetic field in the outflow, then polarimetry of starlight shining through the outflow can measure the projected magnetic field geometry. If the pitch angle of a spiral magnetic field originating in the YSO accretion disk produces a very tight spiral, then the measured field geometry will be perpendicular to the flow \citep{hod89}. Polarimetry at infrared wavelengths was used by \citet{jon03} and \citet{ito99} to study the magnetic field geometry in the massive molecular outflow associated with DR21. Most of their observations were consistent with a magnetic field pointing radially outward, aligned with the axis of the outflow.

A similar conclusion was reached by \citet{hod90} using optical polarimetry of stars in the vicinity of Cepheus A. However, optical polarimetry can not penetrate deep into obscured regions, and is more likely to measure the larger scale magnetic field geometry in which the star formation region is embedded. Infrared (IR) polarimetry has the advantage of being able to penetrate the considerable dust extinction that typically accompanies deeply embedded massive YSOs \citep{hod89}. This allows us to probe lines of sight more likely to pass through the denser portions of the outflow. Imaging polarimetry provides additional information on the reflection nebulosity produced by the bipolar outflow. Furthermore, at IR wavelengths, reflection nebulosity in heavily obscured regions close to the driving stellar source can be probed.

Polarimetry at longer wavelengths where the aligned dust grains are in emission avoids the need for background field stars. However, mm- and sub-mm polarimetry of regions with one or more massive outflows are sensitive to the magnetic field geometry associated only with warm dust near the massive YSO's, not the more distant dust in the bipolar outflow. \citet{gre97} found a rough correlation between the position angle of the magnetic field in the warm dust centered on the YSO and the direction of the outflow. When the outflow lies in the plane of the sky, the magnetic field tends to be perpendicular to the outflow axis, just the opposite of what is found further out in the outflow \citep{jon03, ito99}. 

In this paper we employ the technique of imaging polarimetry at $H~(1.65~\micron)$ and $K~(2.2~\micron)$ of three compact regions with massive star formation: G$192.16-3.82$, W42, and Cepheus A. G192.16 and Ceph A are well known outflow systems and Ceph A has had optical and near-IR polarimetry of surrounding stars \citep{hod89,hod90}. W42 was not previously known to contain an outflow, but we have discovered a very reddened, elongated reflection nebula in the cluster indicative of a bipolar outflow. For all three systems we present IR polarimetry of stars within and nearby the central star forming region and imaging polarimetry of the reflection nebulosity associated with the outflow.

\section{OBSERVATIONS}

All observations were made with NSFCAM \citep{shu94} in polarimetry mode on the NASA IRTF 3-m telescope at a plate scale of 0\farcs3 per pixel, resulting in a field of view of $77\arcsec$ square. In polarimetry mode NSFCAM utilizes a rotating half-wave plate at the entrance window of the camera and a wire grid polarizer in the secondary filter wheel. This allows the observer to form images at polarization position angles of 0, 45, 90 and 135$^\circ$ on the sky. From these we compute Stokes quantities $Q$ ($I_{0}-I_{90}$) and $U$ ($I_{45}-I_{135}$) for stars and nebulosity within the image. Details of the NSFCAM (+Polarimeter) technique are given in \citet{jon97}, \citet{jon00}, and \citet{kel04}. Table 1 summarizes observational details and parameters of our targets.

Photometric and instrumental polarization calibrations were achieved by observing stars from the list of \citet{eli82}. Unreddened early type stars in this list were assumed to have no polarization at the 0.1\% level. Measurements of several of these stars indicate that the instrumental polarization is less than 0.1\%. Because our technique requires a finite time between waveplate positions, sky and transmission fluctuations create systematic errors for single measurements at the $\pm 0.2\%$ level. These systematic effects are the primary source of error for bright objects. Position angle and polarization efficiency calibrations were made by observing S1 in $\rho$ Oph, which was assumed to have P$_{H} = 3.9\%$, $\theta_{H} = 28\degr$ and P$_{K} = 1.95\%$, $\theta_{K} = 28\degr$. Additional checks on the position angle callibration were provided by observing BN in Orion and AFGL 2591. Photometric magnitudes were transfromed from the natural IRTF system \citep{tok02} to the modern MKO system using the transformation relations described in \cite{kel04}.

Our ability to measure the polarization for stars depends on three factors. If the star is very faint, it will be impossible to measure the fractional polarization, given the typical values of a few percent for interstellar polarization. If the star is very bright, but has a very low intrinsic polarization, it will also be impossible to measure the polarization. Finally, stars with images confused with nearby stars, stars in the negative (sky) beam, or stars embedded in bright nebulosity will also be impossible to accurately measure. In this paper we report polarimetry of stars only if the signal-to-noise for the fractional polarization $(P/\sigma_P)$ is greater than 3.

\section{DISCUSSION}

\subsection{Grain Alignment}

Interstellar polarization has long been attributed to the alignment of interstellar grains through some interaction with the magnetic field, such as a modified form of the classic Davis-Greenstein mechanism \citep{dav51}. There is little doubt that in the diffuse ISM between stars magnetic alignment of some sort is the principle mechanism at work (for a review of grain alignment, see \citet{jon97b, laz03}). This conclusion is based on the pattern of optical polarization vectors in the Milky Way \citep{hei00} and comparisons between Faraday studies at radio frequencies and optical/IR polarimetry in external galaxies \citep{jon97b}. However, in specific locations not probed by these tests, other mechanisms might be dominant. These include the original mechanical mechanism of \citet{gol52} that used streaming motions of the gas as well as more advanced treatments \citep{laz95} and magneto-mechanical mechanisms such ambipolar diffusion \citep{rob95}.

Although there is no direct evidence for grain alignment by these mechanical mechanisms at this time, one location they might be expected to play a role is in the outflows from massive YSOs, which will certainly have significant streaming motions in the gas. The direction of grain alignment (long axis) for the original Gold mechanism is along the direction of the flow. This is perpendicular to the grain alignment we find based on polarimetry of background stars shining through the outflows discussed below. However, not all alternatives to the Davis-Greenstein mechanism align the grains in this manner. With the caveat that other alignment mechanisms may play a role in massive YSO outflows, in this paper we will assume magnetic grain alignment by some modified version of the Davis-Greenstein mechanism \citep{jon97b}, where the direction of the polarization vector for light in extinction corresponds to the magnetic field direction. 

\subsection{G$192.16 -3.82$}

G192.16 has been extensively studied in CO \citep{sne90,she98}, other molecules \citep{mol96}, maser emission \citep{cod96}, and near-IR imaging \citep{hod94}. Molecular line maps show a single a bipolar outflow from a young star, most likely unconfused with other outflows, extending $1-2\arcmin$ to the East and West at a position angle close to $90\degr$. There is an infrared reflection nebula that coincides with the molecular outflow \citep{hod94}. The outflow is centered on IRAS 05553+1631, which is coincident with a cm- and mm- continuum source \citep{she99, she02}. The radio continuum fluxes are consistent with an early B star at a distance of 2 Kpc \citep{she98}. High angular resolution cm- and mm- continuum observations show evidence for a solar system size accretion disk surrounding the exiting star \citep{she02}. G192.16 has a very long bipolar outflow, about 3 pc, with Herbig-Haro objects extending up to 10 pc from the star \citep{dev99}.

Our $K$-band image of G192.16 core and portions of the outflow to the East and West is shown in Fig.~1. The $K$-band polarization vectors of six stars located along the line of sight are shown in the top panel of Fig.~1. The coordinates and polarization of these stars are listed in Table 2. The cm- and mm- continuum source believed to correspond to the central source of the outflow is located just to the West of the bright reflection nebulae in the central image, marked by a white cross. The white patches in Fig.~1 are due to stars in the negative (sky) beam.

The polarization of stars seen toward G192.16 on the sky is shown in the larger context of the surrounding ISM in Fig.~2, where we include two stars (\#7 and \#8) well off the outflow axis to the South (also listed in Table 2). These two stars have polarization vectors closely aligned with the galactic plane at a position angle of $140^\circ$. For the distance and direction of G192.16, we are looking largely across spiral arms in the Milky Way and would expect the magnetic field to lie predominately in the plane \citep[e.g.][]{jon97b}. Examination of this region of the sky in the large data base of optical polarimetry of stars compiled by \citet{hei00} also shows the interstellar polarization vectors lie in the plane of the galaxy. Thus, G192.16 is embedded in a large-scale magnetic field aligned with the galactic plane. The outflow axis of G192.16 is not aligned with the surrounding magnetic field, but is oriented at a significant angle of $\sim 50\degr$ to the galactic plane.

With one exception (\#3), polarization vectors for the six stars in the same region of the sky as the outflow are not aligned with the interstellar magnetic field. This strongly suggests the interstellar polarization of these stars is influenced by dust associated with G192.16. Stars \#1 and \#6 are $\approx 1\arcmin$ NE and SW of the core respectively and have polarization vectors pointing radially away from the central source. These stars are likely sampling the magnetic field geometry in the outflow, indicating that at a projected distance of  0.6 pc from the embedded YSO the magnetic field is largely radial, as was the case for DR21. However, within a projected distance of a few tenths of a parsec, the polarization vectors are more chaotic. Star \#3 is aligned with the local ISM and may be a foreground object, but the other three stars show no simple pattern of polarization position angles. Unlike the DR21 outflow, where the observations were consistent with a purely radial field geometry everywhere in the outflow, the magnetic field geometry close to the exciting star in G192.16 must be more complex. 

Deep, high resolution imaging of the region surrounding the YSO at 2.1~$\mu$m has been reported by \citet{rem03}. They find a faint 2~$\mu$m source about $1 \arcsec$ East of the radio and mm continuum source. The $K$-band flux from this source shows little evidence for Br$\gamma$ or shocked molecular hydrogen emission. The source is essentially undetected at $H~(1.65~\micron)$, indicating it is very red. \citet{rem03} speculate that this source is either directly viewed photospheric emission from the central star or a bright knot of diffuse emission associated with the accretion disk. Below we examine the imaging polarimetry of this source and try and determine its true nature.

$K$-band polarimetry of the reflection nebulosity associated with the G192.16 outflow is illustrated in the bottom panel of Fig.~1 and listed in Table 3. These measurements were made using a synthetic $4.5\arcsec \times 4.5\arcsec$ aperture with our $I$, $Q$ and $U$ images. All of the polarization vectors are consistent with simple scattering of light from a central illuminating source close to the location of the cm- and mm- continuum source. In Fig.~3 we examine the region of bright nebulosity close to the illuminating source in more detail. The upper left-hand panel in Fig.~3 shows a grey scale image of the inner region. Here we see a bright patch of reflection nebulosity and a fainter source, indicated by the cross lines, about $3\arcsec$ West of the bright peak. This fainter source is the source discovered by \citet{rem03} who had significantly better seeing (0.6\arcsec) than in our image ($1.7\arcsec$). Based on our astrometry using POSS stars in the vicinity of G192.16, the faint source is located at RA = $5^{h} 58^{m} 13\fs6$, Decl. = $+16\degr 31\arcmin 58\arcsec \pm 1\arcsec$ J(2000), in excellent agreement with \citet{rem03}. 

Using a $1.5\arcsec \times 1.5\arcsec$ synthetic aperture we made polarization measurements of the nebulosity shown in Fig.~3. The results of these measurements, uncorrected for seeing effects, are shown in the lower left panel of Fig.~3. The source discovered by \cite{rem03} is highly polarized, and is very likely reflection nebulosity, not the central star itself. Because of seeing effects, flux from the brighter patch of nebulosity can contaminate the polarized intensity we measure for nearby regions. We can roughly subtract this contamination from patches nearby the bright source by using the point spread function of nearby stars to estimate the amount of flux underlying the patch of sky we are trying to measure. For the faint source, contamination from the bright patch was about 30\% of the total intensity. When the correction is made to $I$, $Q$ and $U$, we obtain the polarization vectors in the lower right panel of Fig.~3. These vectors point back to an illuminating source that is $1.5\pm1.0\arcsec$ West of the suspected stellar source. Within our estimated position error of $1\arcsec$, this is also the location of the compact continuum source. We conclude that the massive YSO responsible for the molecular outflow and the illumination of the nebulosity in G192.16 is coincident with the cm- and mm- continuum source and is not visible at $K$.

Our images are sufficiently deep to put new constraints on the extinction in front of the central star. In areas of our images well away from bright nebulosity, our 3~$\sigma$ detection limit is about $K = 18$. Near the position of the exciting star, however, we are limited to about $K = 16.5$ due to the bright underlying nebulosity. We estimate the central star must be fainter than this limit. Using the expected value for the apparent $K$ magnitude of a B2 star at 2 Kpc suggested by \citet{rem03} of $K = 9.7$, we derive $A_{K} \geq 6.8$. Using the extinction law from \citet{rie85}, this corresponds to greater than 65 magnitudes of extinction at V. The infrared color for this amount of extinction is $E(H-K) \geq 4.2$, placing the $H$ band brightness of this star well out of reach of our observations.

\subsection{W42}

W42 (= G$25.4-0.2$S) is a moderately compact, obscured HII region toward the inner Galaxy at $\ell = 25.4^{\circ}$. Continuum radio maps at 5~GHz were made by \citet{woo85} that showed an elongated emission structure with two cores smaller than the 6$\arcsec$ beam. At the 5~Kpc distance they adopted, the radio continuum observations were consistent with an O6.5-7 star as the primary ionizing source. \citet{blu00} have surveyed the stellar content of W42 with imaging and spectroscopy in the near-IR. They find a cluster of stars with a $K$ band luminosity function similar to the Trapezium. The HII region is powered by a single star that they classify as O5-6.5, based on the classification system of \citet{han97}. \citet{blu00} suggest a closer distance of $\sim 3$ kpc, based on spectroscopic parallax arguments applied to the O star and its dereddened $K$ magnitude \citep[see also][]{les85}. The spectrum of other bright stars in the cluster are featureless, suggesting they are still YSO's.

Our $K$-band image of W42 is shown in Fig.~4 along with the polarization vectors of fourteen stars located along a line of sight to the HII region. The coordinates and polarization data for these stars are listed in Table 4. Note that the polarization vectors are relatively uniform in position angle. There is no evidence for more than one component to the magnetic field sampled along the line of sight to W42. The stars within $30\arcsec$ of the exciting O star (star \#12 in Table 4, star \#1 in \citet{blu00}) have a mean position angle of 18$^\circ$ with a dispersion of 12$^\circ$. The stars to the East and West of the central cluster have a position angle that lies in the plane of the Milky Way at about 28$^\circ$. This is also the position angle of the long axis of the continuum radio emission observed by \citet{woo85}.

\citet{woo85} speculate that the elongated radio emission may be indicative of a bipolar source. Although there is no bipolar nebula clearly visible in our $K$-band image nor in the images in \citet{blu00}, close examination of the polarimetry does indeed reveal a bipolar nebula. In the left panel of Fig.~5 we show a grey scale image of the inner $30\arcsec$ of the HII region in total intensity in the $K$ band. In the right panel we show an image in polarized intensity ($I_P = \sqrt{Q^2 + U^2}$) with the polarization vectors for five regions of highly polarized nebulosity superimposed. These five regions were measured using a synthetic aperture $2.7\arcsec\times2.7\arcsec$ in size. Highly polarized reflection nebulosity, almost unrecognizable in the total intensity image, is clearly visible in the polarized intensity image. The polarization vectors are consistent with an illuminating source located between the two brightest blobs of nebulosity, indicated by the cross marks on the right and bottom of the polarized intensity image. If there is a luminous YSO located between the two bright blobs of reflection nebulosity, it would have coordinates RA = $18^{h} 38^{m} 14\fs 6$, Decl. = $-6\degr 48\arcmin 00\arcsec$ (J2000). The long axis of the reflection nebulosity is oriented at about 15$^\circ$, essentially the same orientation as the magnetic field through the entire cluster.

The newly discovered bipolar outflow is very heavily reddened by extinction. The $H-K$ color of the two brightest blobs is $H-K \sim +2.4$, redder than any of the surrounding stars \citep[Fig.~4 in][]{blu00}. Since reflection nebulosity can be intrinsically much bluer than $H-K = 0.0$ (depending on the spectral energy distribution of the illuminating star), our measured color translates to a visual extinction of $A_V \ga 55$. The illuminating star is presumably behind additional dust associated with an equatorial accretion disk, explaining why it is not visible at $K$. \citet{blu00} argue that we see the association and its primary O star on the front side of a dense, molecular cloud. The newly discovered bipolar nebula is clearly deeper into this denser region than the cluster, indicating ongoing star formation just behind the interface between the HII region and the molecular cloud. This reminds us of the Trapezium, which has a comparable luminosity and a similar geometry with respect to our line of sight from the Earth. 

Optical polarimetry of stars in the surrounding sky taken from the compilation of \citet{hei00} clearly shows a general magnetic field direction in the diffuse ISM surrounding W42 that lies in the plane of the Galaxy at $PA=28\degr$. The magnetic field threading W42 is close to this position angle. We are presented with a picture of a very uniform magnetic field geometry threading through the entire cluster, and linking up smoothly with the field in the surrounding diffuse ISM. The newly discovered bipolar nebula has a long axis aligned in the same direction as the threaded magnetic field, indicating that the polar axis of the illuminating star is also aligned with the local magnetic field. This is in contrast to G192.16 and DR21, where the axis of the bipolar nebula and the magnetic field in the outflow is well away from alignment with the galactic plane.

\subsection{Ceph A}

Cepheus A East contains a well studied, relatively nearby molecular outflow driven by one or more massive, very young stars. At a distance of 725 pc \citep{bla59} the Far-IR luminosity is about $2.4 \times 10^{4} L_{\odot}$, which is consistent with several early B stars. The bipolar outflow is quite complex \citep{tor93, nar96} and is suggestive of multiple outflow sources. The molecular line observations have also been interpreted as clumps in two sides of a poorly collimated bubble powered by one or more YSOs \citep{cor93}.

Radio continuum observations at a resolution of $1\arcsec$ \citep{hug84} reveal several compact sources, which they associate with early B stars. Source number 2 in their map, commonly referred to as HW2 in the subsequent literature, is usually considered to be the primary source of the outflow. This source was undetected at $K$ by \citet{cas96}, who present imaging polarimetric observations of Ceph A on a $2.7\arcsec$ grid at $J$, $H$ and $K$. Analysis of the highly polarized reflection nebulosity surrounding HW2 allowed \citet{cas96} to show that the origin of the scattered light from dust in the outflow is primarily from HW2. They also found that within $10-15\arcsec$ of HW2, the polarization vector pattern was more complex than simple single scattering and clearly required multiple scattering. 

\citet{goe98} made a detailed study of Ceph A East using broad and narrow band imaging from $1.4 - 5~\micron$. They find two regions of shock-excited line emission, one $20\arcsec$ to the NE of HW2 and one $20\arcsec$ to the ESE of HW2. They identify HW3 (not HW2) as the primary driving source of the shock emission to the ESE and consequently favor a multiple outflow model. They find that the reflection nebulosity to the NE of HW2 and HW3 is exceptionally red in their $K-L'$ image, indicating it is behind significant extinction. \citet{goe98} were unable to detect HW2 at $L'$ or $M$, implying $A_V > 200$ if it is a bare B1 star. In their schematic diagram of the Ceph A region, \citet{goe98} associate the rotating core observed in CS line emission by \citet{nar96} with a 'Fluffy Torus' centered on HW2 and oriented at a position angle of about $135^{\circ}$.

\citet{gle99} measured the polarization toward HW2 at 1.3~mm and found the magnetic field was oriented at $170^\circ$, which is roughly perpendicular to the outflow axis. This result is consistent with the conclusion of \citet{gre97}, who argue that when the bipolar outflow lies in the plane of the sky, the magnetic field observed at the Far-IR/sub-mm peak is perpendicular to the outflow axis. \citet{hod90} observed I band polarimetry of nearby stars and found the magnetic field geometry $1\arcmin - 4\arcmin$ away from HW2 was inconsistent with a single component. Unlike G192.16 and W42, the optical polarimetry of stars within a few degrees of Ceph A does not present a clear picture of the surrounding interstellar magnetic field. Using the data base of interstellar polarimetry compiled by \citet{hei00}, we plot the polarization vectors for stars within a few degrees of Ceph A in Fig.~7. South of the star forming region the polarization vectors are nicely aligned with the galactic plane, indicating the interstellar magnetic field lies in the plane along that line of sight, as expected. However, $1^\circ$ North and West, some stars show interstellar polarization vectors that lie in the plane of the Galaxy, while others lie at a position angle of about $125^\circ$. We will argue below that the foreground interstellar magnetic field is aligned with the plane of the Milky Way disk ($PA = 65^\circ$), the magnetic field threading the Ceph A complex is almost perpendicular to the plane ($PA = 125^\circ$) and that the magnetic field in the outflow is radial with respect to HW2/3.

A false color image of Ceph A using our $H$ and $K$ band images is shown in Fig.~6. The location of HW2 is shown as a green cross. The reddest tones correspond to $H-K = +4.0$ and the bluest tones correspond to $H-K = +1.75$, which is still quite red. Several stars are bluer than $H-K = +1.75$ and they are saturated at a blue color. Depending on the intrinsic colors of the illuminating source, the $H-K$ color of reflection nebulosity would usually be significantly bluer than any of the values we measure in Ceph A. The entire reflection nebulosity must be behind significant interstellar extinction, very likely associated with the star forming region itself. This intervening extinction is optically thickest close to the central source, and becomes thinner as one moves North and East from the massive YSO. A similar conclusion was reached by \citet{goe98} from their $K-L'$ image.

In Fig.~8 we plot a number of features seen in Cepheus A over an underlaying grey scale of the $K$ band intensity. The solid contours are from the radio continuum map of \citet{hug84}. Their source 2 (HW2) is associated with the primary central massive YSO that powers and illuminates the outflow. The narrow rectangles are the polarization vectors of the reflection nebulosity on a $2\arcsec$ grid. The solid black lines are the $H$-band polarization vectors of the five stars listed in Table 6. The position angle of the galactic plane is $65^\circ$, very close to the position angle of star \#4. 

Based on the $H-K$ colors we observe for the nebulosity, the extinction within $10-15\arcsec$ of HW2 is too great for background stars to shine through \citep[see][]{goe98}. Star \#1 in Table 6 is outside the bright nebulosity and is relatively red. It has a position angle very close to the position angles measured by \citet{hod89} in $H$ band for stars a few arc minutes north and northwest of Star \#1. We hypothesize this star is either embedded within the Ceph A complex or less likely, shining through from behind. This would imply that the interstellar magnetic field in the immediate vicinity of Ceph A is at a position angle of $\sim125^\circ$ \citep{hod89}.

Star \#2 is much bluer than Star \#1 and has a similar position angle, but is still outside the bright nebulosity and may also be sampling aligned grains in Ceph A. Star \#3 is much bluer than the underlying reflection nebulosity and can not be shining through the outflow. Polarimetry of Star \#3 is sampling the magnetic field geometry well in front of the outflow and is consistent with a foreground field lying in the plane of the Galaxy. Stars \#4 and \#5 are only slightly bluer than the underlying nebulosity and are probably sampling aligned grains associated with the near side of the Ceph A outflow. Polarimetry of Star \#4 could be interpreted in two ways. It could be dominated by the foreground extinction, causing the position angle to lie in the plane. But, the position angle points directly back to HW2/3 and could be indicative of a magnetic field in the outflow pointing radially to the exciting B stars. The position angle of star 5 also points directly back to HW2/3 and is most likely sampling the magnetic field in the outflow as well. Note that \citet{hod90} and \citet{hod89} measured  several stars $1-2\arcmin$ East of Star \#5, all of which have position angles very similar to Star \#5. We argue that the magnetic field in the outflow at a distance greater than $20\arcsec$ from the cluster of massive YSOs, is radial.

A polarization map at a spatial resolution of $0.9\arcsec$ is shown in Fig 9. The polarization vectors for the reflection nebulosity are dominated by single scattering directly from the central sources at distances of more than $5-10\arcsec$ from HW2/3. The polarization vectors to the North and West of the embedded stars are consistent with scattering of light only from  HW2. The polarization vectors to the NE and E, however, are illuminated from a location about 5$\arcsec$ SE of HW2. We do not see evidence for this at a coarser grid spacing in the polarization map of \citet{cas96}. Our polarization map requires more than one source of light being scattered by dust in the Ceph A outflow. Dust in the outflow to the N and W is illuminated only by the YSO associated with HW2. The dust in the outflow to the NE and E is seeing at least one additional source to the SE of HW2, perhaps source HW3. \citet{goe98} associate the shocked gas coincident with HW7 in the radio map with outflow from HW3, not HW2. Note that both the axis of the elongated radio emission in HW7 and the polarization vector for star 5 point directly back to HW3. It may be that HW3 has brightened at near-IR wavelengths since the observations of \citet{cas96} and is now contributing more to the reflection nebulosity.

The polarization vectors strongly depart from simple centro-symmetric scattering along a band stretching across HW2 from the SE to the NW at a position angle of $\sim 135^{\circ}$. This morphology corresponds well in position angle to the smaller 'Fluffy Torus' described in \citet{goe98} and the rotating gas described by \citet{nar96}. \citet{cas96} conclude that this departure from centro-symmetric scattering as likely due to multiple scattering in dust near the central source. We are probably looking through an optically thick disk of material extending at least $15\arcsec$ (0.05 pc) to the NW and SE of the central cluster of B stars. Light passing through this region will be scattered at least twice and loose the simple centro-symmetric polarization pattern produced by single scattering.

Based on all of the available observations of interstellar polarization toward Cepheus A, we propose the schematic picture of the magnetic field geometry within and in front of Ceph A shown in Figure 10. The foreground (and possibly background) magnetic field is closely aligned with the galactic plane at a PA of $\sim 65^{\circ}$. The field threading the overall star forming complex is at a very different orientation of $\sim 125^{\circ}$. At the Far-IR/sub-mm peak, essentially at the location of HW2 and HW3, the magnetic field is roughly perpendicular to the outflow at $\sim 170^{\circ}$. We say roughly, because there are probably multiple outflows and the reflection nebulosity presents a very broad fan-like structure on the sky. Note, however, that the field direction at the Far-IR/sub-mm peak is perpendicular to the outflow axis in the model proposed by \citet{nar96}. Finally, the interstellar polarization to the East of the core cluster, as measured by Hodapp and ourselves, shows the magnetic field well out in the outflow is radial.

\section{CONCLUSIONS}

We have made imaging polarization observations of three regions of massive star formation at near-IR wavelengths. These observations allowed us to measure the interstellar polarization of stars along the line of sight to these regions and to measure the polarization of reflection nebulosity associated with embedded YSOs. Both G192 and Ceph A are well known bipolar molecular outflow sources with well identified obscured massive YSOs driving the flows and illuminating the reflection nebulosity. In W42 we have discovered a previously unknown bipolar nebula and identified a probable location for the central star responsible for the observed reflection nebulosity. 

\subsection{Reflection nebulosity}

The reflection nebulosity in G192.16 was bright enough in several locations for us to measure the polarization due to scattering of light from the central YSO by dust in the outflow. The observed polarization is entirely consistent with illumination by a single stellar source. This position of this object is coincident with the well studied point-like cm- and mm- continuum source previously identified as the illuminating YSO. The YSO is undetected in our maps at $2~\micron$ down to $K=+16.5 \;(3\sigma)$. 

In Ceph A, the polarization map of the reflection nebulosity shows evidence for more than one source of illumination in addition to HW2, possibly HW3. Neither of these sources are detected at $2~\micron$ in our maps. The polarization in the reflection nebulosity close to HW2 and stretching from HW2 for several arcseconds perpendicular to the outflow shows clear evidence of multiple scattering. This region probably contains the outer extremes of a large accretion disk surrounding HW2. 

\subsection{The magnetic field geometry}

In W42 and Ceph A, the main axis of the outflow and the local magnetic field are within 15$^\circ$ of the galactic plane. In G192.16, the main axis of the outflow and the magnetic field in the outflow are misaligned by at least 60$^\circ$ with respect to the galactic plane. In W42 the magnetic field threading the HII region is well aligned with the galactic plane, and shows no evidence for additional components to the magnetic field geometry. The magnetic field geometry within Ceph A is complicated, but we believe the field threading the complex is not aligned with the galactic plane, even though the outflow axis may be. In G192.16, the magnetic field geometry threading the compact HII region could not be determined.

In all three regions, the local magnetic field is aligned with the direction of the outflow at distances of several tenths of a parsec and further from the central source. In G192.16 the magnetic field geometry closer than a few tenths of a parsec to the central source is probably chaotic, and does not exhibit a radial geometry. In Ceph A, we have no observations of the magnetic field geometry within $20 \arcsec$ of the exciting star(s), but sub-mm observations indicate the field at the location of the embedded YSO(s) is perpendicular to the outflow axis. In W42, the newly discovered bipolar nebula is too small in angular extent on the sky for our technique to be able explore the embedded magnetic field geometry. The axis of the nebula is, however, exactly aligned with the magnetic field in the HII region and the galactic plane.

The magnetically driven model for YSO outflows predicts a spiral magnetic field in the core of the outflow along the outflow axis. Our observations do not rule this model out if the field is a tight spiral only at small distances of less than $\sim 0.2$~pc from the star. At greater distances, our observations are consistent with a purely radial field. The tendency for the magnetic field at the location of the embedded YSO(s) (measured by sub-mm polarimetry) to lie perpendicular to the flow is probably due to a toroidal field in the extended equatorial disk, not in the outflow itself \citep[see the various model geometries discussed in][]{gre97}. Neither our technique of measuring interstellar polarization of background stars nor sub-mm polarimetry with large beam sizes will be able to probe the magnetic field geometry in the outflow close to, but separated from, the embedded star and its accretion disk. High spatial resolution far-IR and mm- polarimetry will be necessary.

\acknowledgements
TJJ, CEW and MSK acknowledge support from the National Science Foundation Grant AST-0205814.  The authors also wish to thank the day-crew and staff of the NASA IRTF (especially Lars Bergknut) for their assistance at the telescope.

\newpage

 FOR HIGH QUALITY FIGURES VISIT http://www.astro.umn.edu/\~~tjj/ \\
CLICK ON ~~~~~ \textbf{Outflow Paper Figures}

\figcaption[tjjones.fig1.eps]{Grey scale image of G$192.16-3.84$ at $K ~(2.2~\micron)$. The location of the mm- point source is indicated by a white cross. White objects in the frames are due to stars in the sky beam. The top panel has the polarization vectors for six of the stars listed in Table 2 superimposed on the image. The bottome panel has the polarization vectors for the reflection nebulosity measured at different locations in the outflow and listed in Table 3. All of the polarization vectors are consistent with light scattered from a single source located at the position of the mm- continuum source.
\label{fig1}} 

\figcaption[tjjones.fig2.eps]{Polarization of stars nearby and south of G$192.16-3.84$. The cross mark indicates the location of the mm- continuum source identified as the YSO powering the outflow. The molecular outflow is aligned roughly East-West. The galactic plane lays at a position angle of 120$^\circ$. Note that the position angles of the two stars to the South (Table 2) show that the surrounding interstellar magnetic field lays in the galactic plane. However, the stars close to G192.16 are clearly at very different position angles than the magnetic field in the surrounding diffuse ISM. 
\label{fig2}} 

\figcaption[tjjones.fig3.eps]{Detailed view of the reflection nebulosity near the central source in G$192.16-3.84$. The upper left panel shows a grey scale image at $2.2~\micron$ of the nebulosity. The cross marks indicate the location of the faint source discovered by \citet{rem03}. The thick solid line locates a cut through the image plotted in the upper right hand panel. This illustrates how flux from the bright peak will contaminate the fainter sources nearby due to seeing effects. The bottom left hand panel shows the polarization vectors at various locations in the region uncorrected for contamination by flux from the bright peak. The lower right hand panel shows these same polarization vectors after a rough correction for contamination from the bright peak. These vectors are consistent with a source of illumination $1.5\pm1.0\arcsec$ west of the source discovered by \citet{rem03}. 
\label{fig3}} 

\figcaption[tjjones.fig4.eps]{Grey scale image of W42 (= G$25.4-0.25$) at $K~(2.2~\micron)$. Superimposed on the image are the polarization vectors for the stars listed in Table 4. The bright star in the center of the cluster (star 12 in Table 4) was identified by \citet{blu00} as an O5-O6.5 star and the primary ionizing source in the HII region. The surrounding interstellar magnetic field is closely aligned with the galactic plane.
\label{fig4}} 

\figcaption[tjjones.fig5.eps]{The left panel shows a grey scale image in total intensity at $K$ of the inner 30$\arcsec$ of W42. The bright star at center left is the O star identified by \citet{blu00}. The right panel shows the same region, but in {\it polarized} intensity. Note the highly polarized reflection nebulosity seen in the polarized intensity image is barely visible in the total intensity image. The polarization vectors for six locations in this reflection nebulosity are superimposed and the coordinates and polarimetry are listed in Table 5. The two arrows in the right hand panel indicate the location of a possible illuminating star for this reflection nebulosity.
\label{fig5}} 

\figcaption[tjjones.fig6.eps]{False color image of the Cepheus A outflow from $H$ and $K$ band images. The green cross indicates the location of the primary illuminating source. The reddest tones correspond to $H-K = +4.0$. The bluest tones correspond to $H-K = 1.75$. Stars with $H-K$ colors bluer than +1.75 are saturated at the bluest color. The reflection nebulosity is clearly much more heavily reddened near the central star.
\label{fig6}}


\figcaption[tjjones.fig7.eps]{Optical polarimetry of stars within a few degrees of Cepheus A (marked with a cross) taken from the compilation of \citet{hei00}. The galactic plane is indicated by the dashed line.
\label{fig7}} 

\figcaption[tjjones.fig8.eps]{Grey scale image of Ceph A at $K$. The solid contours are the radio continuum observations of \citet{hug84}. Their Source 2 (HW2) is associated with the central star powering and illuminating the outflow. The thin rectangles are the polarization vectors of the nebulosity measured on a 2$\arcsec$ grid spacing. The polarization vectors of the five stars in Table 6 are plotted as solid black lines. 
\label{fig8}} 

\figcaption[tjjones.fig9.eps]{Polarization vectors of the reflection nebulosity in Ceph A plotted on a 0.9$\arcsec$ grid. The underlying intensity contours at $K$ are shown to aid in location. The coordinates of HW2 are shown by a cross.
\label{fig9}} 

\figcaption[tjjones.fig10.eps]{Schematic diagram of the magnetic field geometry in Cepheus A, based on the interstellar polarization of stars and sub-mm polarimetry centered on the central cluster of massive YSOs.
\label{fig10}}

\begin{deluxetable}{llcc}
\tablenum{1}
\tablewidth{0pt}
\tablecaption{Observing Log}
\tablehead{
\colhead{Object} & \colhead{Date} & \colhead{Filters} & \colhead{Mode}}
\startdata
G192.16 Core, East &Feb. 14, 2002 & $K$ & Polarimetry \\
G192.16 West & Feb. 15, 2002 & $K$ & Polarimetry \\
W42 & July 10, 2003 & $H$, $K$ & Polarimetry \\
Cepheus A & July 9, 2003 & $K$ & Polarimetry \\
Cepheus A & July 10, 2003 & $H$ & Polarimetry \\
\enddata
\end{deluxetable}

\begin{deluxetable}{cccrrr}
\tablenum{2}
\tablewidth{0pt}
\tablecaption{Stars in G192.16 -3.84}
\tablehead{
\colhead{Star} & \multicolumn{2}{c}{RA J2000 Decl.} & \colhead{$P_K (\%)$} & \colhead{$\epsilon_P (\%)$} & \colhead{$\theta_K (^\circ)$}\\
\colhead{} & \colhead{$5^h 58^m$} & \colhead{$+16^\circ$} & \multicolumn{3}{c}{}}
\startdata
1 & 19\fs3 & 32\arcmin $30\arcsec$ & 1.6 & 0.3 & 52 \\
2 & 15\fs6 & 32\arcmin $19\arcsec$ & 2.3 & 0.3 & 93 \\
3 & 14\fs6 & 31\arcmin $55\arcsec$ & 1.5 & 0.3 & 155 \\
4 & 15\fs2 & 31\arcmin $39\arcsec$ & 1.3 & 0.3 & 87 \\
5 & 15\fs7 & 31\arcmin $37\arcsec$ & 1.9 & 0.3 & 35 \\
6 & 11\fs2 & 31\arcmin $21\arcsec$ & 2.6 & 0.8 & 178 \\
7 & 14\fs3 & 30\arcmin $00\arcsec$ & 1.8 & 0.4 & 141 \\
8 & 10\fs0 & 30\arcmin $04\arcsec$ & 2.2 & 0.5 & 138 \\
\enddata
\end{deluxetable}

\begin{deluxetable}{cccrr}
\tablenum{3}
\tablewidth{0pt}
\tablecaption{Reflection Nebulosity in G192.16 -3.84}
\tablehead{
\colhead{Patch} & \multicolumn{2}{c}{RA J2000 Decl.} & \colhead{$P_K (\%)$} & \colhead{$\theta_K (^\circ)$} \\
\colhead{} & \colhead{$5^h 58^m$} & \colhead{$+16^\circ$} & \multicolumn{2}{c}{}}
\startdata
1 & 18\fs4 & 31\arcmin $51\arcsec$ & 45 & 12 \\
2 & 16\fs1 & 31\arcmin $49\arcsec$ & 35 & 12 \\
3 & 15\fs4 & 31\arcmin $51\arcsec$ & 39 & 15 \\
4 & 13\fs8 & 31\arcmin $57\arcsec$ & 37 & 11 \\
5 & 12\fs9 & 31\arcmin $53\arcsec$ & 18 & 165 \\
6 & 12\fs3 & 32\arcmin $15\arcsec$ & 31 & 46 \\
7 & 12\fs3 & 32\arcmin $00\arcsec$ & 17 & 7 \\
8 & 11\fs8 & 31\arcmin $39\arcsec$ & 62 & 138 \\
\enddata
\end{deluxetable}

\begin{deluxetable}{cccrrr}
\tablenum{4}
\tablewidth{0pt}
\tablecaption{Stars in W42}
\tablehead{
\colhead{Star} & \multicolumn{2}{c}{RA J2000 Decl.} & \colhead{$P_K (\%)$} & \colhead{$\epsilon_P (\%)$} & \colhead{$\theta_K (^\circ)$} \\
\colhead{} & \colhead{$18^h 38^m$} & \colhead{$-6^\circ$} & \multicolumn{3}{c}{}}
\startdata
1 & 17\fs6 & 47\arcmin $21\arcsec$ & 0.9 & 0.3 & 32 \\
2 & 16\fs3 & 47\arcmin $15\arcsec$ & 2.2 & 0.3 & 5 \\
3 & 17\fs7 & 47\arcmin $54\arcsec$ & 5.1 & 0.9 & 41 \\
4 & 16\fs2 & 47\arcmin $45\arcsec$ & 2.2 & 0.3 & 24 \\
5 & 15\fs8 & 47\arcmin $45\arcsec$ & 2.1 & 0.3 & 178 \\
6 & 15\fs4 & 47\arcmin $39\arcsec$ & 0.8 & 0.8 & 28 \\
7 & 14\fs9 & 47\arcmin $41\arcsec$ & 2.3 & 0.4 & 17 \\
8 & 14\fs7 & 47\arcmin $44\arcsec$ & 1.3 & 0.5 & 10 \\
9 & 13\fs6 & 47\arcmin $48\arcsec$ & 1.8 & 0.3 & 18 \\
10 & 14\fs5 & 47\arcmin $52\arcsec$ & 0.8 & 0.3 & 160 \\
11 & 15\fs3 & 47\arcmin $51\arcsec$ & 2.2 & 0.3 & 8 \\
12 & 15\fs3 & 47\arcmin $58\arcsec$ & 2.1 & 0.3 & 13 \\
13 & 14\fs6 & 48\arcmin $05\arcsec$ & 0.6 & 0.3 & - \\
14 & 15\fs3 & 48\arcmin $06\arcsec$ & 1.8 & 0.3 & 10 \\
15 & 15\fs6 & 48\arcmin $13\arcsec$ & 2.1 & 0.3 & 178 \\
16 & 13\fs5 & 48\arcmin $15\arcsec$ & 1.8 & 0.6 & 19 \\
\enddata
\end{deluxetable}

\begin{deluxetable}{cccrr}
\tablenum{5}
\tablewidth{0pt}
\tablecaption{Reflection Nebulosity in W42}
\tablehead{
\colhead{Patch} & \multicolumn{2}{c}{RA J2000 Decl.} & \colhead{$P_K (\%)$} & \colhead{$\theta_K (^\circ)$}\\
\colhead{} & \colhead{$18^h 38^m$} & \colhead{$-6^\circ$} & \multicolumn{2}{c}{}}
\startdata
1 & 14\fs69 & 47\arcmin $54\arcsec$ & 13 & 100 \\
2 & 14\fs66 & 47\arcmin $57\arcsec$ & 32 & 97 \\
3 & 14\fs55 & 48\arcmin $02\arcsec$ & 20 & 97 \\
4 & 14\fs50 & 48\arcmin $06\arcsec$ & 20 & 106 \\
5 & 14\fs47 & 48\arcmin $11\arcsec$ & 26 & 101 \\
\enddata
\end{deluxetable}

\begin{deluxetable}{ccccrrr}
\tablenum{6}
\tablewidth{0pt}
\tablecaption{Stars in Cepheus A}
\tablehead{
\colhead{Star} & \multicolumn{2}{c}{RA J2000 Decl.} & \colhead{$H-K$} & \colhead{$P_H (\%)$} & \colhead{$\epsilon_P (\%)$} & \colhead{$\theta_H (^\circ)$}\\
\colhead{} & \colhead{$22^h 56^m$} & \colhead{$+61^\circ$} & \multicolumn{4}{c}{}}
\startdata
1 & 16\fs0 & 02\arcmin $19\arcsec$ & 2.75 & 5.9 & 0.2 & 123 \\
2 & 17\fs3 & 02\arcmin $24\arcsec$ & 0.73 & 0.9 & 0.2 & 154 \\
3 & 19\fs8 & 02\arcmin $21\arcsec$ & 0.61 & 0.8 & 0.2 & 54 \\
4 & 22\fs5 & 02\arcmin $01\arcsec$ & 1.39 & 4.9 & 0.3 & 63 \\
5 & 22\fs6 & 01\arcmin $42\arcsec$ & 1.45 & 2.1 & 0.3 & 94 \\
\enddata
\end{deluxetable}

\end{document}